\documentstyle[preprint,aps]{revtex}  %% Draft Style

\begin{document}
\flushbottom
\draft

\title{\bf  Z=110--111 Elements and the Stability of Heavy and Superheavy
  Elements}

\author { Cheng-Li  Wu,$^{1} $ Mike  Guidry,$^{2,3}$ and Da Hsuan Feng$^4$}

\address{
$^{(1)}$Department of Physics, Chung Yuan Christian University, Chung-Li,
  Taiwan 32023 ROC \\
$^{(2)}$Department of Physics and Astronomy, University of Tennessee,
  Knoxville, Tennessee 37996 \\
$^{(3)}$Physics Division, Oak Ridge National Laboratory,
  Oak Ridge, Tennessee 37831 \\
$^{(4)}$Department of Physics and Atmospheric Sciences, Drexel University,
  Philadelphia, PA 19104
  }

\date{\today}

\maketitle

\begin{abstract}
  The recent discovery of isotopes with Z=110--111 suggests evidence for
(1) a monopole--monopole interaction that does not appear explicitly in
Nilsson--Strutinsky mass systematics, and (2) a competition between
$SU(2)$ and $SU(3)$ dynamical symmetries that has been predicted for this
region.  Our calculations suggest that these new isotopes are near spherical,
and may represent a true island of superheavy nuclei, but shifted downward
in neutron number by these new physical effects.
\end{abstract}


\narrowtext

  The search of superheavy elements is a long-term goal of nuclear
structure physics \cite{greiner}.
In the 1960s a series of calculations using the liquid drop plus shell
correction approach to determining nuclear masses and energy surfaces suggested
the existence of a group of relatively stable nuclei that would be separated in
neutron and proton number from the known heavy elements by a region of much
higher instability \cite{nil69}.  This group of nuclei come to be known as the
{\em island of superheavy elements,} and its conjectured existence provided a
major initial justification for a generation of heavy ion accelerators and
experiments performed on those accelerators. To this date, no convincing
evidence for the existence of this island has been  presented, despite many
accelerator-based experiments and many searches for evidence in nature of such
elements.

  Meanwhile, in a series of difficult experiments performed in the past
  decade \cite{mun81,mun82,mun84}, evidence has accumulated for
extension of the known elements to larger proton number, culminating in
the recent discovery of several isotopes having proton number
$Z=110-111$ \cite{hof95}.  The usual view is that although these
isotopes are very heavy indeed, they are {\em not} examples of the
originally sought island of superheavy elements.  One reason is that
the recently-discovered isotopes of elements 110 and 111  have neutron
numbers approximately 20 units lighter than the predicted superheavy
island. Thus, in this view these new elements represent the tail of
the distribution of ``normal'' elements, and the predicted superheavy
elements represent a qualitatively different set of nuclides.

  In 1987 a new approach to nuclear masses based on the Fermion
  Dynamical Symmetric Model (FDSM) was introduced \cite{wu87}. This was
subsequently refined and extended to a systematic calculation of masses
for heavy and superheavy elements (Z=82--126, and
N=126--184)\cite{han92,wu94b}.  There is good agreement between the
masses predicted by this theory and the experimentally measured ones.
For example, the r.~m.~s.\ error for all available masses above $Z=82$
and $N=126$ is approximately 0.22 MeV. In Table  1, we show the masses
of the heaviest elements and compare with a variety of calculations for
these masses.  We note that the FDSM calculations were not tuned
specifically to these heaviest elements.

One of the most interesting features of the FDSM
calculations\cite{han92} is that one finds an island of superheavy
elements with considerably lower neutron number than that of
traditional calculations.  The center of the island is around $Z=114$
and $N=164$; this shifted island
 is also found to be more stable in the FDSM calculations than the
original island in traditional calculations,  and we find that these
nuclides correspond to nuclei having near-spherical shapes.  The shell
correction associated with this minimum is illustrated in Fig.\
\ref{esurface}a. The location of the new predicted maximal shell
stabilization, the traditional location of the island of superheavy
stability, and the location of the most stable isotopes for each
elementt, including new isotopes of elements 110--111, are also shown
in Fig.\ \ref{esurface}a. According to the FDSM prediction, these newly
discovered superheavy elements are at the edge of the superheavy
island, and the commonly expected superheavy minimum near the N=184
neutron closed shell does not even exist.

Furthermore, from the FDSM
mass calculation one can obtain the $\beta$ stability and fission
stability lines. The $\beta$ stability line is defined as the line
connecting nuclides with maximum binding energy per nucleon for a given
$A$,  while the fission stability line is defined by connecting the
nuclides  with maximum binding energy per nucleon for a given Z.
They are found to be
$ A=Z/0.390$ and $ A=Z/0.409$, respectively.  As can be seen from
Fig.\ \ref{esurface}b, these two predicted stability lines agree with
data.  The most stable isotopes (the longest lifetime)
lie on the predicted $\beta$ stability line when
spontaneous fission is not important (Z $\leq$ 100), and then quickly
switch to the fission stability line.  For Z$>$106, the
experimentally discovered elements, including the newly discovered $Z=110$
$(A=269)$ and $Z=111$ $(A=272)$ all lie on the predicted fission stability
line.  The trends for the most stable nuclei are also shown in
Fig.\ \ref{esurface}a.

The above prediction is based on the commonly used Woods--Saxon
single-particle potential. As was mentioned in ref.~\cite{han92}, by
arbitrarily altering the  s.~p.\ levels one could shift the superheavy
minimum to a position which is near the previous predictions.
Therefore, until one has complete confidence about the
s.~p.\ potential, predicting the precise location of the superheavy island is
uncertain. Nevertheless, it is important to understand the physical reason for
two FDSM predictions that differ fundamentally from distorted mean field
(DMF) predictions: (1)~Why do spherical nuclei
exist near midshell for neutrons in the FDSM? (2)~Why within the FDSM framework
does a  shell correction minimum not seem to exist at $N=184$?

Although both the FDSM and the DMF invoke shell correction concepts, they
differ in two fundamental
aspects.  (1)~The Hamiltonians of the two approaches are
extremely different.  In the DMF, the Hamiltonian is the sum of BCS
deformed s.~ p.\ energies. The nuclear mass is the sum of the deformed
liquid drop mass (with pairing) and the deformed s.~ p.\ shell
correction. In the FDSM, the Hamiltonian is the sum of {\em spherical}
s.\ p.\ energies and two-body interactions (which include monopole
pairing-plus-quadrupole, quadrupole-pairing, and monopole--monopole
interactions). The nuclear mass is equal to the sum of the {\em spherical}
liquid drop mass (without pairing) plus the spherical s.~p.\ shell correction
and the expectation value of the two-body interactions. Thus, the pairing
correlation and quadrupole effects (nuclear deformation) in the FDSM are
treated through the two-body correlations. (2)~The shell corrections in the two
approaches are quite different. The DMF shell correction is defined as the
fluctuating part of the deformed s.~p.\ energies (i.~e.\ the difference between
the sum of the deformed s.~p.\ energies and a smooth part), with the
calculation carried out using the semiempirical recipe developed by Strutinski.
 In the FDSM, the shell correction is the sum of two parts: a spherical
single-particle shell correction and the expectation value of the two-body
residue interactions.  The latter goes beyond the (deformed) mean field, while
the former shares the same spirit as the DMF.  In the FDSM, the simple Fermi
Gas Model (not the Strutinski recipe) is used to describe the smooth part of
the spherical s.~p.\ energies.  It is well known that the Fermi Gas Model is
inadequate within the usual shell correction procedure, but in the FDSM is
appears to be adequate to give a very good description of masses.

It is of particular interest for this problem to investigate the role played by
the various two-body interactions.  Indeed, as we shall show, the difference of
the FDSM predictions from those of the DMF method are direct consequences of
the interactions. (1)~The subtle competition between pairing and quadrupole
correlations together with the Pauli effect can create a small window near
midshell for very heavy nuclides to be spherical or near spherical. (2)~The
monopole--monopole interaction, which is increasingly repulsive as the mass
increases, tends to wash out the shell correction minimum near N=184.

Let us first address the stabilization of spherical shapes near midshell.
In Fig.~2, we show the competition between pairing and quadrupole correlations
in the $Z=82-126$ and $N=126-184$ shells.
In the FDSM, pairing gives an $SU(2)$ dynamical symmetry and
hence spherical shape. Quadrupole interactions, on the other hand,
favor an $SU(3)$ dynamical symmetry and hence
deformed shapes \cite{wu94b}. For $SU(3)$ dynamical symmetry
the completely symmetric $SU(3)$ representation is
energetically most favorable.  Therefore in the absence of additional
physical constraints, this representation will
be the system`s ground state for deformed nuclei
(see the dashed line of Fig.~2a). The spherical vibrational mode,
(corresponding in the FDSM to $SU(2)$ dynamical symmetry)
is usually dominant when the particle number is small;
hence it will usually
occur for nuclides near closed shells.

However, such considerations fail to account fully for the role
of the Pauli effect in limiting collectivity \cite{fen88,gui93c,wu94b}.
In particular,  the most energetically favorable $SU(3)$ irrep is
forbidden by the Pauli principle when the  particle (hole) number
in the normal-parity levels exceeds one third of the shell (which
correspond to $99\leq Z \leq 116$ and $152 \leq N \leq 170$ for the shells we
are considering in the present problem) \cite{fen88,gui93c,wu94b}.  This is why
the solid line of Fig.~2a (corresponding to the $SU(3)$ curve of
$<V_{pq}>$) has a W shape, thus allowing
the $SU(2)$ symmetry to compete favorably near midshell for the heaviest
nuclei. Within the FDSM, this is the physics of the
narrow window of spherical or near-spherical stability
in which the predicted superheavy minimum
near $Z=114$ and $N=164$ appears to lie \cite{han92}.

In Fig.~\ref{su2su3}a, the remaining shell corrections $\langle
M_{sh}^{0}\rangle$
are shown. The sum of $\langle M_{sh}^{0}\rangle$ and $\langle V_{pq}\rangle$,
the total shell correction, is displayed in Fig.~\ref{su2su3}b. Notice that the
inclusion of $\langle M_{sh}^{0}\rangle$ does not significantly alter the
competition between $SU(2)$ and $SU(3)$, thus allowing the most favorable
states beyond $Z=108$ to possess $SU(2)$ symmetry (open circle line) and the
maximum negative shell correction to be around $Z=114$ ($N=165$).

However, $\langle M_{sh}^{0} \rangle$ causes the
shell correction  to increase dramatically at the end of the shell
because $\langle M_{sh}^{0} \rangle$ contains the
spherical s.~p.\ shell correction
as well as the monopole--monopole interaction.
The s.~p.\ shell correction has a $\cap$ shape with positive shell corrections
near mid-shell and large negative corrections at both ends.  The
monopole--monopole interaction $V_{mono}$ is approximated by a quadratic
function of the number of particles:
\begin{equation}
V_{\rm\scriptsize mono}= a_{\alpha}
+b_{\alpha}N_{p}+c_{\alpha}N_{p}^{2}+d_{\alpha}N_{n}
+e_{\alpha}N_{n}^{2}+f_{\alpha}N_{p}N_{n}
\label{Eq.1}
\end{equation}
The parameters of this equation are determined from fitting to known masses
to be:
for $\alpha = SU2$,\\
\begin{equation}
\begin{array}{llrllrllr}
a_{\alpha} & = & -13.75, \hspace{0.1in}& b_{\alpha}& = &-4.572,\hspace{0.1in} &
c_{\alpha}& = &0.4293  \\
d_{\alpha}& = &-4.889, \hspace{0.1in} &e_{\alpha}& = &0.3306, \hspace{0.1in} &
f_{\alpha}& = &-0.2915\\
\end{array}
\label{Eq.2}
\end{equation}
for  $\alpha = SU3$,\\
\begin{equation}
\begin{array}{llrllrllr}
a_{\alpha}& = &-5.570, \hspace{0.1in} &b_{\alpha}& = &-5.790, \hspace{0.1in} &
c_{\alpha}& = &0.3713  \\
d_{\alpha}& = &-6.806, \hspace{0.1in} &e_{\alpha}& = &0.3587, \hspace{0.1in} &
f_{\alpha}& = &-0.1095\\
\end{array}
\label{Eq.3}
\end{equation}
This interaction has the opposite
effect of the  s.~p.\ shell correction.  If there
were no monopole--monopole interaction included in our calculation,
our shell correction would have the commonly expected $\cap$ shape,
with a large and positive value near midshell and
maximum negative value at shell closure.
The negative linear terms ($b_{\alpha}$ and $d_{\alpha}$), and
the n--p interactions ($f_{\alpha}$) in $V_{\rm\scriptsize mono}$,
will partially cancel the
positive s.~p.\ shell correction, thus reducing the value of
$\langle M_{sh}^{o}\rangle$
near midshell.  The repulsiveness of the monopole--monopole interactions
between like particles ($c_{\alpha}$ and $ e_{\alpha} >0$) will
cause the shell correction at the end of the shell to increase rapidly.
This is why the FDSM predicts a large negative shell correction
at closed shells, but only for
very heavy nuclei.  For lighter ones (e.~g.\ rare earths)
the shell corrections will continue to have
the $\cap$ shape, because the valence like-particle number is not large enough
so that the monopole--monopole interaction can cancel the large negative
s.~p.\ shell correction at the end of the shell.

The behavior of the rapid increase at the end of the shell in actinide
region ($Z=82-126$, $N=126-184$) is
a unique property of the form of the FDSM shell correction.
In the DMF, the predicted mass shell corrections always have the
$\cap$ shape with negative values at the shell closure.   This
difference will lead to  very different predictions
about the unknown superheavy
elements, although in the known regions
the two approaches give similar results.
This difference may be understood as follows.
The DMF employs a liquid drop mass formula, and for a given deformed s.~p.\
energy scheme the Strutinsky recipe is used to extract the fluctuating
contribution to the mass.
Since the s.~p.\ energy scheme is empirical
and thus strongly influenced by the known data, it is not surprising that
there are no dramatic changes when extrapolated to the unknown regions.
However,  there is no obvious physical guidance to access the accuracy of the
shell corrections computed in this way.
For the FDSM mass formula, the rapid increase at the end of the shell for the
actinide shell correction
 is a direct consequence of the monopole--monopole interaction, which should be
present on general shell model grounds.
Although the linear monopole--monopole interaction terms may already
be included in the deformed s.~p.\ energy scheme, the quadratic
terms definitely have not.

Next we shall present a simple study of the alpha decay energies and the
corresponding decay lifetimes for the heaviest elements. Our calculations are
based on the simplest one-dimensional barrier penetration model. We compare our
results employing the masses determined in Ref.\ \cite{han92} with observations
for the very heavy elements in Fig.~\ref{qalpha}. We find that despite the
simplicity of the barrier penetration portion of the calculations, there is
quite reasonable agreement with the experimental mass as well as alpha-decay
energies of the new isotopes, especially since our results for the masses were
reported prior to the measurements and were based on an {\em analytical model}
that was applied to all masses with $Z> 82$ and $N> 126$. In particular, there
was no attempt to optimize results for the heaviest elements. We note that the
predicted alpha decay energies exhibit the correct qualitative trends (e.~g., a
maximum at $Z=109$).  The absolute half-lives are off by several orders of
magnitude for $Z=110-111$, but the trends are correct (e.\ g. a minimum at
$Z=109$), and the error is not large considering the exponential nature of the
barrier penetration process and the crude model employed for alpha decay.

The FDSM predictions concerning the masses of the heaviest elements have some
testable consequences, though the required experiments are formidable.
(1)~There should exist  nuclides of elements $Z=$ 112--114 that are as stable
as those of elements 110--111 (see Fig.~\ref{esurface} and Fig.~\ref{qalpha}).
(2)~Beyond $Z \approx 116$ , the heavy nuclei should rapidly become less stable
(see  Fig.~\ref{qalpha}), and the region of traditional superheavy nuclei
should be quite unstable (see   Fig.~\ref{esurface}a) if the commonly used
Woods--Saxon s.~p.\  spectrum is reliable. (3)~The nuclides in this new region
of superheavy nuclei ($Z \approx 110-114$ and $N \approx 160-165$) are expected
to be either spherical or very deformation soft because of the $SU(2)$ minimum
that competes  favorably with the $SU(3)$ minimum as a consequence of the
dynamical Pauli effect (see Fig.~\ref{su2su3}); this structure should influence
properties such as the alpha decay systematics. (4)~The implied shift of the
r-process path closer to the stability valley should also have observable
consequences, but this may require an improved understanding of the
astrophysical environment for the r-process.

In summary,  the  heaviest isotopes yet observed, corresponding to elements
having proton number 110--111, have masses that were predicted by principles of
dynamical symmetry to a rather high precision.  We have suggested that the
success of this prediction provides support for a previously-conjectured
 monopole--monopole component in the mass equation that becomes increasingly
important in very heavy nuclei, and for a successful competition of an $SU(2)$
dynamical symmetry with the $SU(3)$ dynamical symmetry as a consequence of
the dynamical Pauli effect of the Fermion Dynamical Symmetry Model. As a
result of this $SU(2)$ symmetry, we expect the newly-discovered isotopes
of elements 110--111 to be nearly spherical or very deformation soft.
Furthermore, we suggest that elements 110--111 may represent, not just the
heaviest isotopes yet discovered, but the first examples of the
originally-conjectured island of superheavy nuclei with neutron number
approximately 15--20 units smaller than in traditional calculations because of
the effects discussed above.  Conversely, we predict no stable nuclides in the
vicinity of the traditional superheavy island if the Woods-Saxon s.~p.\
spectrum is reliable. These conjectures could be tested by further observations
of isotopes in the $Z=110-114$ region, by a continued failure to find
superheavy elements at their historically expected location, and by observable
associated with r-process element production.

Nuclear physics research at CYCU is supported by the National Science Council.
Theoretical nuclear physics research at the University of Tennessee is
supported by the U.~S. Department of Energy through Contract No. \
DE--FG05--93ER40770. Oak Ridge National Laboratory is managed by Martin
Marietta Energy Systems, Inc.\ for the U.~S. Department of Energy under
Contract No.\ DE--AC05--84OR21400. Nuclear physics research at Drexel
University is supported by the National Science Foundation.

%\bibliographystyle  {prsty}
%%--> Physical Review Bib style

%\bibliography{sd}          %%  Read bibliography references and
                                      %%  order from  file sd.bib

\begin{onecolumn}

%-------------------------------------
% Table style with no lines between rows
%-------------------------------------

\begin{table}[tbh]
\begin{center}
Table 1.  Experimental and Theoretical Shell Corrections in
MeV for the Heaviest Elements.
\\
\begin{tabular}{lllllllll}
%\hline
$^{256}104$&
$^{258}105$&
$^{260}106$&
$^{262}107$&
$^{264}108$&
$^{266}109$&
$^{269}110$&
$^{272}111$&
Reference
\\
\hline
%\hline
94.25(03)&
101.84(34)&
106.60(04)&
114.68(38)&
119.82(30)&
128.39(35)&
134.80(32)*&
141.70(37)*&
Exp \cite{aud93,hof95}
\\
\hline
94.15&
101.98&
106.8&
114.82&
120.02&
128.16&
133.31&
140.49&
FDSM \cite{han92}
\\
95.90&
103.41&
106.87&
116.10&
121.28&
129.44&
&
144.83&
Myers \cite{ato76}
\\
94.84&
102.22&
105.81&
115.00&
120.40&
128.43&
&
144.04&
Groote \cite{ato76}
\\
95.6&
102.6&
106.8&
114.9&
120.4&
127.8&
&
142.6&
Seeger \cite{ato76}
\\
94.37&
101.64&
105.68&
114.78&
120.27&
128.44&
&
143.82&
Liran \cite{ato76}
\\
95.77&
103.01&
108.13&
115.50&
&
&
&
&
M\"{o}ller \cite{mol81}
\\
95.69&
103.11&
108.12&
115.71&
121.09&
129.04&
135.51&
143.44&
M\"{o}ller \cite{mol88}
\\
93.78&
101.00&
105.81&
113.18&
118.34&
126.06&
132.39&
140.18&
M\"{o}ller \cite{mol88}
\\
93.38&
100.97&
105.73&
113.45&
118.73&
126.65&
133.08&
140.93&
M\"{o}ller \cite{mol95}
\\
94.52&
&
107.04&
&
120.47&
&
&
&
Patyk \cite{pat91}
\\
94.38&
&
106.93&
&
120.47&
&
135.46&
&
Cwiok \cite{cwi94}
\end{tabular}
\end{center}
* Masses for $^{269}110$ and $^{272}111$ are extracted from the Q values of
$\alpha$-decay $^{269}110\longrightarrow$ $^{265}108$ and
$^{272}111\longrightarrow$ $ ^{268}109$, respectively \cite{hof95}.
\end{table}

\centerline{\bf Figure Captions}

\begin{figure}[h]
\caption{\protect\label{esurface} (a) Mass shell correction for heavy and
superheavy elements using the dynamical symmetry methods of
Ref.~\protect\cite{han92}. The locations of the most stable nuclei for each
observed element are also indicated by circles with a shadowed trace to guide
the eye. (b) The $\beta$ stability line and the fission stability line.  The
open circles are the most stable nuclei  as indicated in part (a).  }

\end{figure}

\begin{figure}
\caption{\protect\label{su2su3} (a) Competition between $SU(3)$ energy
$V_{\mbox{pq}}(su3)$ and $SU(2)$ energy $V_{\mbox{pq}}(su2)$.  The dashed line
is the $SU(3)$ energy when the dynamical Pauli effect (DPE) is ignored.
$M_{\mbox{sh}}^{0}(su3)$ and $M_{\mbox{sh}}^{0}(su2)$ are mainly the sum of the
spherical s.~p.\ shell correction and the monopole--monopole interactions (see
the text).  (b) The total mass shell corrections for $SU(2)$,  $SU(3)$, and
$SU(3)$ with no DPE, $M_{\mbox{sh}}=M_{\mbox{sh}}^{0}+V_{\mbox{pq}}$. The plot
is along the fission stability line,  $A=Z/0.409$.  In the $V_{pq}$ plots, the
even--odd  pairing difference is ignored  to make the figure more legible, but
in the total mass shell correction plots (b) the even--odd pairing difference
is included.}

\end{figure}

\begin{figure}
\caption{\protect\label{qalpha} Alpha decay energies and alpha decay
half-lives.   Data are represented by dots and the predictions of
Ref.~\protect\cite{han92} are illustrated by the solid lines.}
\end{figure}

\end{onecolumn}
\end{document}